\documentclass[sigconf]{acmart}
\usepackage{enumitem}
\usepackage{makecell}
\AtBeginDocument{%
  \providecommand\BibTeX{{%
    \normalfont B\kern-0.5em{\scshape i\kern-0.25em b}\kern-0.8em\TeX}}}

\copyrightyear{2022}
\acmYear{2022}
\setcopyright{acmlicensed}\acmConference[CAIN'22]{1st Conference on AI Engineering - Software Engineering for AI}{May 16--24, 2022}{Pittsburgh, PA, USA}
\acmBooktitle{1st Conference on AI Engineering - Software Engineering for AI (CAIN'22), May 16--24, 2022, Pittsburgh, PA, USA}
\acmPrice{15.00}
\acmDOI{10.1145/3522664.3528620}
\acmISBN{978-1-4503-9275-4/22/05}

\acmConference[CAIN’22]{CAIN’22 - 1st Conference on AI Engineering – Software Engineering for AI}{May 21-22 2022}{Pittsburgh, PA, USA}
\usepackage{tcolorbox}
\usepackage{tikz}
\usetikzlibrary{shapes.callouts, shadows}
\tikzset{author comment/.style={draw, fill=white, thick, drop shadow}}
\newboolean{CommentON}
\setboolean{CommentON}{true}
\newcommand{\MyComment}[3]{%
	\ifthenelse{\boolean{CommentON}}{%
		\raisebox{-.5ex}
		{\tikz
			\node[x=1ex, y=1ex, inner sep=.5ex,
			rectangle callout,
			callout pointer width=.7ex,
			callout relative pointer={(1.5,-0)},
			author comment]
			{\footnotesize\textsf{#1}};}~%
		\textsf{[}\,\textcolor{#2}{#3}\,\textsf{]}%\xspace
	}{} %else
}

\usepackage{listings}

\newcommand{\lstbg}[3][0pt]{{\fboxsep#1\colorbox{#2}{\strut #3}}}
\lstdefinelanguage{diff}{
  basicstyle=\ttfamily\small,
  morecomment=[f][\lstbg{red!20}]-,
  morecomment=[f][\lstbg{green!20}]+,
  morecomment=[f][\textit]{@@},
}

\usepackage{fontawesome}
\newcommand{\codeSmell}[3]{
  \begin{sloppypar}
  \begin{description}[leftmargin=0em]
    \item[Context] #1
    \item[Problem] #2
    \item[Solution] #3
  \end{description}
  \end{sloppypar}
}

\usepackage{xurl}
\usepackage{hyphenat}
\begin{document}

\title{Code Smells for Machine Learning Applications}
% does your machine learning code smell?

\author{Haiyin Zhang}
\email{haiyin.zhang@ing.com}
\affiliation{%
  \institution{AI for Fintech Research, ING}
  \city{Amsterdam}
  \country{Netherlands}
}

\author{Lu\'{i}s Cruz}
\email{L.Cruz@tudelft.nl}
\affiliation{%
  \institution{Delft University of Technology}
  \city{Delft}
  \country{Netherlands}
}

\author{Arie van Deursen}
\email{Arie.vanDeursen@tudelft.nl}
\affiliation{%
  \institution{Delft University of Technology}
  \city{Delft}
  \country{Netherlands}
}

\renewcommand{\shortauthors}{Zhang, et al.}

\begin{abstract}
The popularity of machine learning has wildly expanded in recent years. Machine learning techniques have been heatedly studied in academia and applied in the industry to create business value. However, there is a lack of guidelines for code quality in machine learning applications. In particular, code smells have rarely been studied in this domain. Although machine learning code is usually integrated as a small part of an overarching system, it usually plays an important role in its core functionality. Hence ensuring code quality is quintessential to avoid issues in the long run. This paper proposes and identifies a list of 22 machine learning-specific code smells collected from various sources, including papers, grey literature, GitHub commits, and Stack Overflow posts. We pinpoint each smell with a description of its context, potential issues in the long run, and proposed solutions. In addition, we link them to their respective pipeline stage and the evidence from both academic and grey literature. The code smell catalog helps data scientists and developers produce and maintain high-quality machine learning application code. 
\end{abstract}




% % A "teaser" image appears between the author and affiliation
% % information and the body of the document, and typically spans the
% % page.

\maketitle

\section{Introduction}

Despite the large increase in the popularity of machine learning applications~\cite{gonzalez-mlgithub2020}, there are several concerns regarding the quality control and the inevitable technical debt growing in these systems~\cite{Sculley2015HiddenTD}. Moreover, machine learning teams tend to be very heterogeneous, having experts from different disciplines that are not necessarily aware of Software Engineering (SE) practices backgrounds and there is a limited number of training and guidelines on machine learning-related software development issues. Hence, software engineering best practices are often overlooked when developing machine learning applications~\cite{simmons2020large,lenarduzzi2021software}. Yet, previous research shows that practitioners are eager to learn more about engineering best practices for their machine learning applications~\cite{haakman2020ai}.

There has been a lot of interest in various machine learning system artifacts, including models and data. Researchers make efforts to improve machine learning model quality~\cite{kohavi1995study} and data quality~\cite{hynes2017data}. However, the quality assurance of machine learning code has not been highlighted~\cite{lenarduzzi2021software}. Recent work studied the code quality for machine learning applications in a general way, finding some code quality issues such as duplicated code~\cite{van2021prevalence} and violations of traditional naming convention \cite{simmons2020large}. These works highlighted the fact that the existing code conventions do not necessarily fit the context of machine learning applications. For example, the typical math notation in data science tasks clashes with the naming conventions of Python~\cite{van2021prevalence}. Thus, we argue that more research is needed to accommodate the particularities of data-oriented codebases.

As an important artifact in the machine learning application, the quality of the code is essential. Low-quality code can lead to catastrophic consequences. In the meantime, different from traditional software, machine learning code quality is more challenging to evaluate and control. Low-quality code can lead to silent pitfalls that exist somewhere that affect the software quality, which takes a lot of time and effort to discover~\cite{zhang2018empirical}. Therefore, it is non-trivial to improve the code quality during the development process and consider code quality assurance in the deployment process.

A common strategy to improve code quality is eliminating code smells and anti-patterns. When we talk about code smells in this paper, we refer them to the pitfalls that we can inspect at the code level but not at the data or model level. We use the term "pitfall" to represent issues that degrade the software quality. Listing \ref{lstling} shows an example of such pitfalls using Python and the Pandas library. In the red (-) part of the listing, an inefficient loop is created. A better alternative is highlighted in green (+), using Pandas built-in API to replace the loop, which operates faster. While some alternative solutions might lead to improvements in runtime efficiency, other solutions might be essential to prevent problems in the long run. For example, previous work shows that code smells affect the maintainability, understandability, and complexity of software~\cite{lacerda2020code}.

\begin{lstlisting}[language=diff, caption=Coding Pitfall Example from~\cite{haakman2020studying}, label=lstling]
import pandas as pd
df = pd.DataFrame([1, 2, 3])

- result = []
- for index, row in df.iterrows();
-   result.append(row[0] + 1)
- result = pd.DataFrame(result)
+ result = df.add(1)

\end{lstlisting}

With the concern of improving machine learning application code quality and easing the machine learning development process, we conduct an empirical study to collect machine learning-specific code smells and provide practical recommendations about the quality in machine learning applications. Thus, we formulate the following research question:  \textit{What are the recurrent code issues that may arise from the peculiarities of machine learning applications?}

The main contributions of this paper are:

1) A catalog of machine learning-specific code smells.

2) A dataset of 1750 papers, 2170 grey literature entries, 87 GitHub commits and 491 Stack Overflow posts for empirical studies.

The replication package for this study is available at \url{https://github.com/Hynn01/ml-smells}. The website with all the smells is published at \url{https://hynn01.github.io/ml-smells/}.

\section{Related Work}
\label{sec:related work}

Code smells are common poor code design choices that negatively affect the systems and violate the best practice or original design vision~\cite{lacerda2020code}. Martin Fowler introduced a general code smell list in his book~\cite{fowler2018refactoring}. Ever since then, code smells have been widely discussed in studies. Many empirical studies have linked code smell proliferation with decreased code quality, increased error proneness, and increased maintainability issues in the long term~\cite{maintainceeffort2013,maintaincodesmells2017,maintainceproblems2013}. 

The prevalence of traditional code smells in machine learning projects was studied in van Oort et al.’s paper~\cite{van2021prevalence}. They ran Pylint on 74 machine learning projects and concluded the most frequent traditioal code smells. Yet, they noted that “the fact that Pylint fails to reliably analyse whether prominent ML libraries are used correctly, provides a major obstacle to the adoption of Continuous Integration (CI) in the development environment of ML systems.” This implies that the context of machine learning applications brings new challenges to the code quality. Therefore, our work addresses machine learning-specific code
smells.
 
Even though there are few code smell studies specific to machine learning application coding, some researchers are studying refactoring and bugs associated with machine learning, which are related to machine learning coding patterns. Most relatedly, Tang et al. studied refactoring in machine learning programs by analyzing 26 projects~\cite{tang2021empirical}. They introduced 14 new machine learning-specific refactorings and seven new machine learning-specific technical debt. However, some of the machine learning programs they analyzed are machine learning tools, while we focus on machine learning applications. We argue that the underlying nature of machine learning libraries and tools is very different from applications. In addition, they focused on classifying different types of refactoring (e.g., "make algorithms more visible"), but they did not extract the code patterns that should trigger such a refactoring. We take a step further by addressing this question. Furthermore, we focus on code smells that cannot be identified by looking at general-purpose smells. For example, while it makes sense to have a type of refactoring for "duplicate model code", its code pattern is no different from the traditional smell "duplicated code". Our paper dives deep into code patterns and examples that are at the machine-learning library API usage level, which is different from their work.

Zhang et al. conducted an empirical study, mining Stack Overflow and GitHub commits to studying the TensorFlow bugs~\cite{zhang2018empirical}. They proposed several bug patterns, which are helpful when debugging deep learning applications. Islam et al. followed up by inspecting Stack Overflow blogs and GitHub commits of five deep learning libraries, including Caffe, Keras, Tensorflow, Theano, and Torch~\cite{islam2019comprehensive}. They adopted some of the root causes of deep learning bugs from ~\cite{zhang2018empirical} and added more root causes. Also, they studied the impacts of bugs, the common patterns of the bugs, and the evolution of the bugs. Humbatova et al. created a comprehensive taxonomy of deep learning bugs by mining GitHub, mining Stack Overflow and interviewing developers~\cite{humbatova2020taxonomy}. The final taxonomy is quite thorough and detailed.

Our work differs from these two studies in four main reasons: 1) we formulate practical coding advice in the form of code smells, to improve the code and avoid potential issues in the long run, 2) we only focus on issues that can be inspected at the code level, 3) we not only focus on pitfalls that lead to potential bugs but also on performance, reproducibility, and maintainability issues, 4) we expand the scope of these smells beyond the deep learning discipline, focusing on other machine learning tasks provided in the libraries Scikit-Learn, Pandas, NumPy and SciPy.

Rajbahadur et al. collected eight data science project pitfalls from a paper and used a model-driven method to detect the pitfalls in the pipeline. Our study differs from theirs by inspecting the faults in the code level to assure the software quality~\cite{rajbahadur2019pitfalls}. Breck et al. learned from the experience with a wide range of production machine learning systems at Google and presented 27 machine learning-specific tests and monitoring needs~\cite{breck2017ml}. However, it does not provide a concrete coding guideline. We go further by building a machine learning-specific code smell catalog and guide machine learning developers towards better coding practices by eliminating code smells.

\section{Methodology}
\label{sec:methodology}

\begin{figure*}[h]
  \centering
  \includegraphics[width=\linewidth]{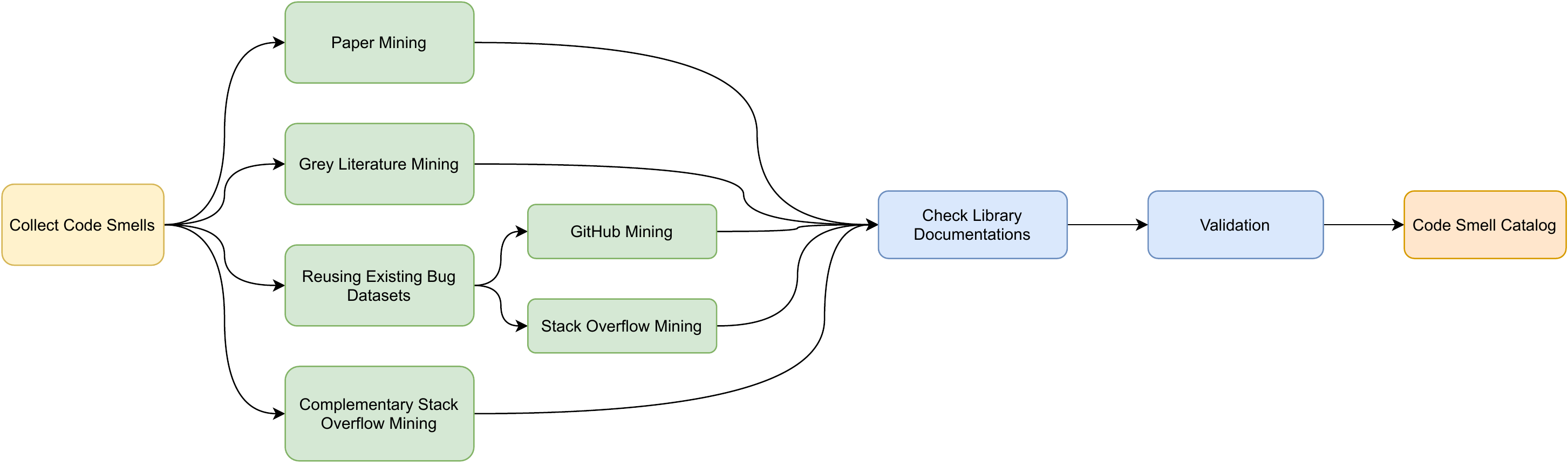}
  \caption{Methodology}
  \label{Methodology}
\end{figure*}

To collect machine learning-specific code smells, we resort to academic literature, grey literature, community-based coding Q\&A platforms (with Stack Overflow), and public software repositories (with GitHub). The general process is depicted in Figure~\ref{Methodology}. We mine papers, grey literature, reuse existing bug datasets, and conduct a complementary Stack Overflow mining. Then we triangulate our collected smells with the recommendations provided in the official documentation of machine learning libraries. In the end, two authors validate the code smell catalog.

\subsection{Paper Mining}
\label{sec:paper_mining}

\begin{figure}[h]
  \centering
  \includegraphics[width=0.7\linewidth]{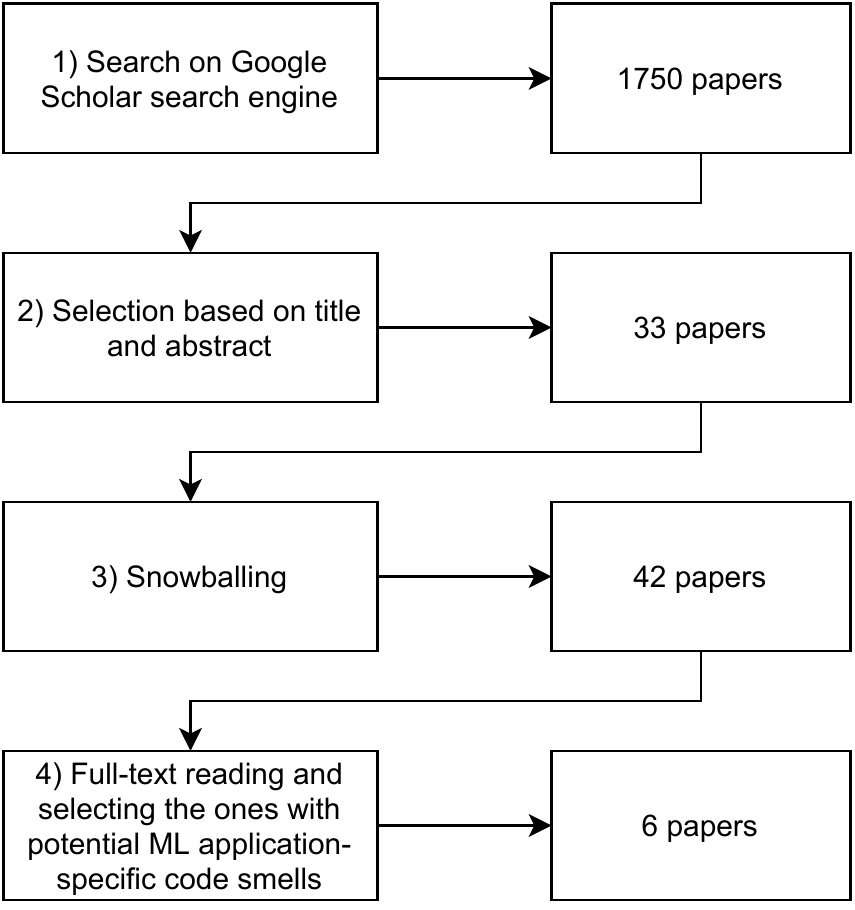}
  \caption{Paper Selection Process}
  \label{paper_selection_process}
\end{figure}

\begin{figure}[h]
  \centering
  \includegraphics[width = \linewidth]{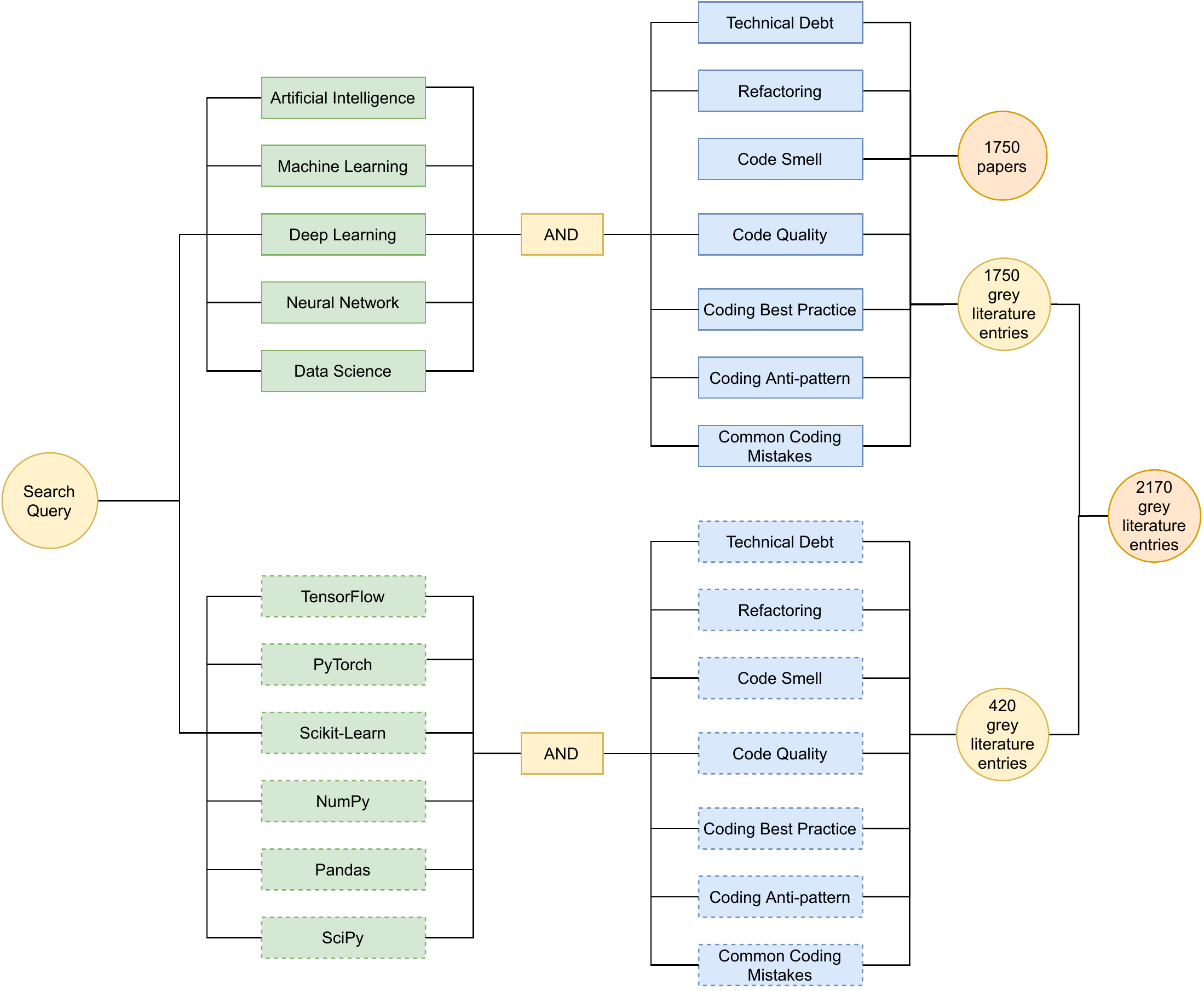}
  \caption{Search Query for Literature and Grey Literature}
  \label{searchquery_literature}
\end{figure}

Our methodology for paper mining is described as follows, and shown as Figure~\ref{paper_selection_process}: 

\textbf{1)  Search on Google Scholar search engine:} To collect papers that potentially contain code smells for machine learning projects, we use terms combining machine learning-related keywords and code quality-related keywords to search. Machine learning-related keywords include ``Artificial Intelligence'', ``Machine Learning'', ``Deep Learning'', ``Neural Network'' and ``Data Science''. Code quality-related keywords include ``Technical Debt'', ``Refactoring'', ``Code Smell'', ``Code Quality'', ``Coding Best Practice'', ``Coding Anti-pattern'' and ``Common Coding Mistakes''. We apply these queries (e.g., ``Machine Learning Technical Debt'') in the Google Scholar search engine, as presented in Figure~\ref{searchquery_literature}. After analyzing papers from the initial result set, we reach a level of saturation for each query after consulting the first five pages of the result. Therefore, we consult the first five pages of the results for each query, i.e., the first 50 results, sorted by relevance at any time by any type. In total, there are 1750 papers ($5\times7\times50$).

\textbf{2) Selection based on title and abstract:} We observe that there are many papers studying machine learning for software engineering (ML4SE) and a few are about software engineering for machine learning (SE4ML). For example, for a paper titled ``Comparing and experimenting machine learning techniques for code smell detection'', we identify it as an ML-for-SE paper and exclude it from our study. When we cannot classify the paper from the title (e.g., ``Toward deep learning software repositories''), we look into the abstract and decide whether to include it in our study. The numbers of selected papers under each query are listed in Table ~\ref{tab:selectedpapers}. After excluding the duplicated ones, our methodology yields 33 papers.

\begin{table*}
  \caption{Number of Selected Papers under Each Query (Duplicates Included)}
  \label{tab:selectedpapers}
  \resizebox{0.8\linewidth}{!}{%
  \begin{tabular}{cccccccc}
    \toprule
    &
    \makecell[c]{Technical\\Debt} &
    Refactoring &
    \makecell[c]{Code\\Smell}  &
    \makecell[c]{Code\\Quality} &
    \makecell[c]{Coding\\Best Practice} &
    \makecell[c]{Coding\\Anti-pattern}  &
    \makecell[c]{Common\\Coding Mistakes} \\
    
    \midrule
    Artificial Intelligence & 7 & 1 & 0 & 2 & 0 & 2 & 0\\
    Machine Learning        & 6 & 1 & 0 & 2 & 7 & 3 & 3\\
    Deep Learning           & 8 & 3 & 0 & 4 & 4 & 1 & 5\\
    Neural Network          & 3 & 0 & 0 & 0 & 0 & 0 & 2\\
    Data Science            & 2 & 0 & 0 & 0 & 2 & 1 & 0\\
  \bottomrule
\end{tabular}}
\end{table*}

\textbf{3) Snowballing:} We apply the forward and backward snowballing method, i.e., browse the reference list of the 33 papers and the list where the paper is cited, select the paper based on the title and abstract as step 2) described, and delete the duplicated papers. We add nine papers after this step.

\textbf{4) Full-text reading and selecting the ones with potential machine learning-specific code smells:} We read the full text of the 42 papers and select the ones with potential machine learning-specific code smells. After this step, we get six final papers. The papers that contribute to the code smell catalog are listed as follows:~\cite{breck2017ml,haakman2020studying,zhang2018empirical,islam2019comprehensive,rajbahadur2019pitfalls,humbatova2020taxonomy}.

\subsection{Grey Literature Mining}
\label{sec:grey_literature_mining}

Many relevant pieces of knowledge about machine learning engineering are being published on the Web by experienced practitioners -- for example, in the format of blog posts. Hence, we use grey literature as a relevant source for machine learning-specific coding advice in this study.

To collect online entries of grey literature, we first resort to the Google search engine with the same queries used above for the research literature (cf. Figure~\ref{searchquery_literature}). We also apply a back-cutting strategy at the end of the fifth page of the result for each query. Hence, there are 1750 entries for this group of search queries, the same number as the paper selection queries. 

Complementarily, we select six machine learning-related Python libraries, namely TensorFlow, PyTorch, Scikit-Learn, Pandas, NumPy, and SciPy, combine them with the code quality-related keywords mentioned in Section~\ref{sec:paper_mining} and form a new group of search queries. Python is widely used for machine learning\footnote{State of Data Science and Machine Learning 2021.\label{footnote:kaggle} \url{https://www.kaggle.com/kaggle-survey-2021}} and the six libraries are the most popular machine learning libraries \footnote{15 Python Libraries for Data Science You Should Know. \url{https://www.dataquest.io/blog/15-python-libraries-for-data-science/}}, covering the two most important steps in machine learning application development -- data processing and model training. For this group of search queries, we reach a level of saturation after analyzing the first result page. Therefore, we consult the first ten results (i.e., first page) the Google search engine provides. There are 420 entries ($6\times7\times10$) for this group of search queries. In total, there are 2170 entries for grey literature mining.

Since not all entries contain actionable coding advice, we select entries by 1) reading the title, 2) reading the first summary, and 3) reading the whole article. Many articles mention some common patterns in machine learning, but most of them are duplicated and are general advice that do not contain code-level pitfalls. 

In the end, we identify eight cornerstone blog posts that contribute to the code smell catalog, as listed in Section~\ref{sec:grey_literature} in the Appendix: \ref{grey:ml_perfect}, \ref{grey:ml_wrong}, \ref{grey:ml-common-mistakes}, \ref{grey:mistakes-by-ds-scientist},  \ref{grey:pytorch_styleguide}, \ref{grey:effective-tensorflow}, \ref{grey:pandas_style_guide}, \ref{grey:sklearn_best_practice}.

\subsection{Reusing Existing Bug Datasets} 

\indent We reuse the dataset provided in the work by Zhang et al. \cite{zhang2018empirical} to mine code smells in Tensorflow applications. Zhang et al. mined the Tensorflow application bugs, analyzed the bugs pattern using 88 Stack Overflow posts as well as 87 GitHub commits and provided a replication package for these bugs (hereinafter called \textit{``TensorFlow Bugs'' replication package}). We reuse their replication package to extract recurrent pitfalls that may generalize to other projects and thus should be documented as code smells.

\subsection{Complementary Stack Overflow Mining} 

\begin{figure}[h]
  \centering
  \includegraphics[width=\linewidth]{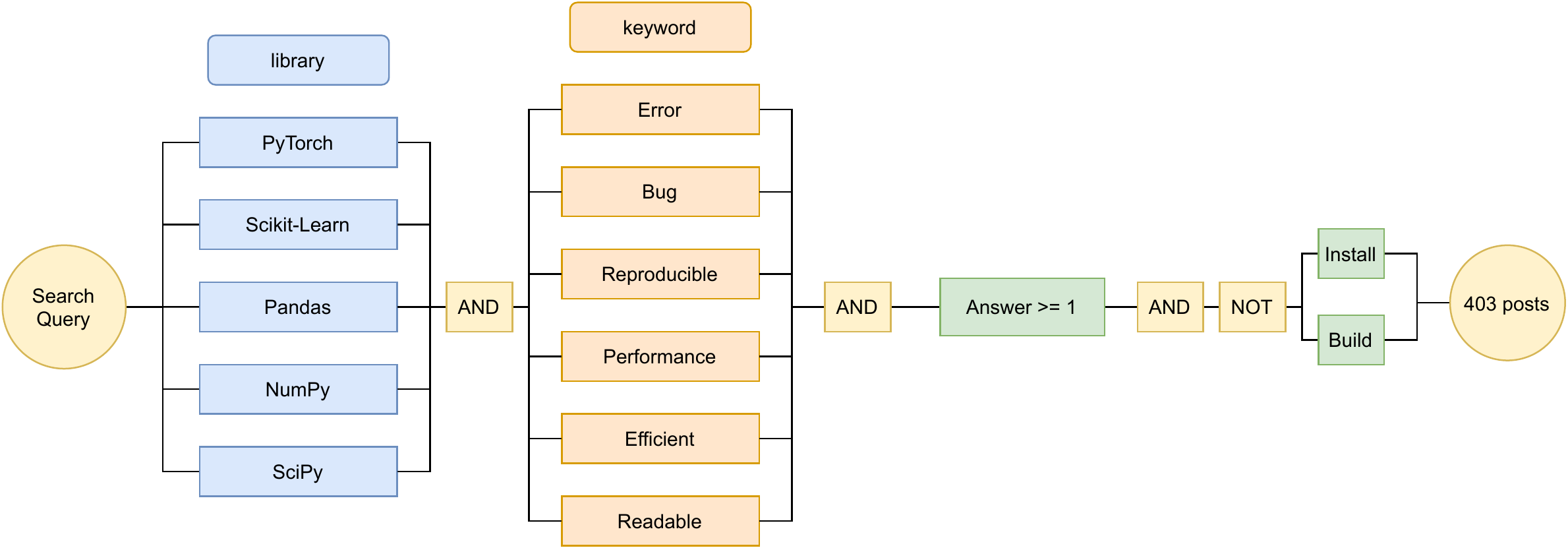}
  \caption{Search Query for Stack Overflow Mining}
  \label{searchquery_so}
\end{figure}

\begin{figure}[h]
  \centering
  \includegraphics[scale=0.6]{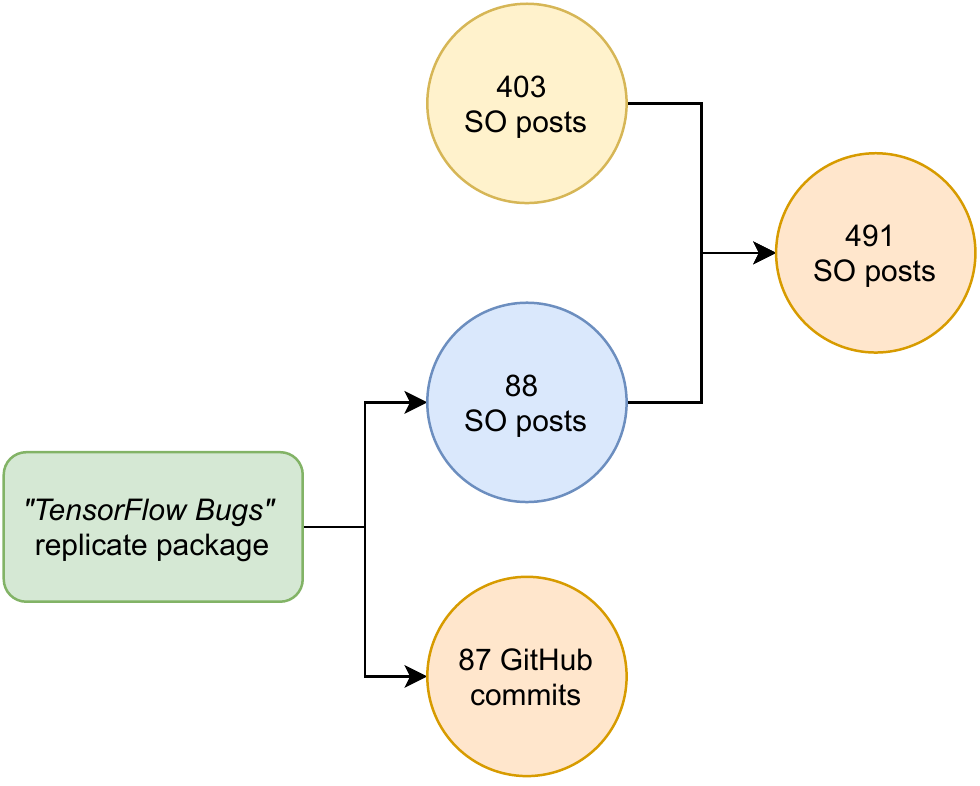}
  \caption{Total Number of Stack Overflow Posts and GitHub Commits}
  \label{so_and_commit}
\end{figure}

\indent After reusing the existing bug datasets, we apply a similar study method to other machine learning libraries. We only check the posts on Stack Overflow at this part without GitHub commits. This is because all the issues have a similar pattern in GitHub and Stack Overflow, as noted by \cite{islam2019comprehensive} and verified in the TensorFlow Bugs replication package~\cite{zhang2018empirical}.

\textbf{1) Library Selection:} We use five libraries: PyTorch, Scikit-Learn, Pandas, NumPy, and SciPy, excluding TensorFlow from the six libraries mentioned in Section \ref{sec:grey_literature_mining}.

\textbf{2) Keyword Selection:} 
To retrieve relevant entries from Stack Overflow we had to redefine our search keywords. We did so because Stack Overflow seldom hosts discussions that directly mention technical debt, smells or refactorings. Entries are mostly related to low-level and straight-to-the-point problems – e.g., performance issues or errors. Hence, keywords had to be adjusted. When collecting posts from Stack Overflow, we use six keywords that are related to typical software quality issues~\cite{international2001iso}: Error, Bug, Reproducible, Performance, Efficient, Readable. 

\textbf{3) Applying Search Terms:} 
We use the notation provided by Stack Overflow to implement the refined queries listed in Figure~\ref{searchquery_so}. It has the following format:

\fbox{\texttt{[{\color{blue}\textbf{library}}] {\textbf{\color{olive}keyword}} \textbf{\color{red}answer:1} \textbf{\color{cyan}-install -build}}} \faSearch{} 

\begin{itemize}
    \item \texttt{{\color{blue}\textbf{library}}} refers to the name of the library we are targeting (e.g., PyTorch, Pandas, etc.).
    \item \texttt{{\textbf{\color{olive}keyword}}} refers to the software quality aspect (e.g., error, bug, etc.).
    \item \texttt{\textbf{\color{red}answer:1}} refers to the entries having at least one answers.
    \item \textbf{\color{cyan}-install -build} these are terms that we exclude from the result set. We filter out \texttt{install} and \texttt{build} because they typically yield results related to configuration and not the codebase.
\end{itemize}

Using the keyword ``Error'' to search gets the maximum number of posts among all libraries. Therefore, we rank all the posts by their votes after applying the term in the search engine, selecting the top 50 posts with the ``Error'' keyword, and selecting the top 10 posts with each of the rest keywords. If the number of posts is fewer than 10, we select all the posts. Then, we delete the duplicated posts for each library. In the end, we get 81, 68, 84, 88, and 82 posts respectively for PyTorch, Scikit-learn, Pandas, NumPy, and SciPy. Together with the \textit{``TensorFlow Bugs'' replication package}~\cite{zhang2018empirical}, we have 87 GitHub commits and 491 Stack Overflow posts in our dataset, as presented in Figure~\ref{so_and_commit}.

\subsection{Validation}
The first author collects all code smells from the empirical study (including paper, grey literature, GitHub and Stack Overflow mining) and discusses the code smell catalog with the second author. We conducted discussion meetings consisting of an introductory discussion of each smell, followed by the analysis of code examples where the code issue had been identified, and the collection of further evidence. We look for references in academic and grey literature that support that particular smell. In total, the first author collected 31 code smells, from which 9 were dropped.

\section{Results}

In this section, we describe 22 machine learning-specific code smells collected from our empirical study. For each smell, we provide a general description followed by the context of the smell, the problem of its occurrence, and the solution. In the end, we summarise all the smells, including the references supporting the smell, the stage of the machine learning pipeline where they are more relevant, and the main effect that arises from having those smells.

We use the notation (n) to cite entries from grey literature, as listed in Appendix~\ref{sec:grey_literature}, where n refers to the nth element in the list.

\begin{table*}[h]
  \caption{Code Smell Catalog}
  \label{tab:codesmellcatalog}
  \resizebox{\linewidth}{!}{%
  \begin{tabular}{llllllll}
    \toprule
    Code Smell & Pipeline Stage & Effect & Type & Literature & \makecell[l]{Grey \\ Literature} & \makecell[l]{GitHub \\ Commits} & SO Posts \\
    \midrule
    
    Unnecessary Iteration  & Data Cleaning &  Efficiency & Generic & \cite{haakman2020studying} & \ref{grey:effective-tensorflow}\ref{grey:pandas_vectorized} & \ref{grey:githubcommit_tf_verctorized} &\\

    NaN Equivalence Comparison Misused & Data Cleaning & Error-prone & Generic & \cite{haakman2020studying} & \\
    
    Chain Indexing & Data Cleaning & \makecell[l]{Error-prone \\ \& Efficiency} & API-Specific: Pandas  & & \ref{grey:pandas_indexing} & &\ref{grey:so_indexing1}\ref{grey:so_indexing2}\\
    
    Columns and DataType Not Explicitly Set  & Data Cleaning & Readability & Generic & & \ref{grey:pandas_style_guide} \\
    
    Empty Column Misinitialization & Data Cleaning  & Robustness & Generic & & \ref{grey:pandas_style_guide}\\  
    
    Merge API Parameter Not Explicitly Set  & Data Cleaning & \makecell[l]{Readability \\ \& Error-prone} & Generic & & \ref{grey:pandas_style_guide}\\   
    
    In-Place APIs Misused  & Data Cleaning &  Error-prone &  Generic & \cite{haakman2020studying} & & \ref{grey:githubcommit_inplace} & \\
    
    Dataframe Conversion API Misused  & Data Cleaning & Error-prone & API-Specific: Pandas & & & & \ref{grey:so_df_conversion} \\
    
    Matrix Multiplication API Misused  & Data Cleaning & Readability & API-Specific: NumPy & &\ref{grey:np_dot} & & \ref{grey:so_npdot}\\
    
    No Scaling before Scaling-Sensitive Operation  & Feature Engineering &  Error-prone &  Generic & & \ref{grey:ml_wrong}\ref{grey:sklearn_scaling} & & \ref{grey:so_scaling}\\
    
    Hyperparameter Not Explicitly Set & Model Training &  \makecell[l]{Error-prone \\ \& Reproducibility} & Generic  & \cite{haakman2020studying}\cite{breck2017ml}\cite{humbatova2020taxonomy}& \\
    
    Memory Not Freed & Model Training &  Memory Issue &  Generic  & \cite{humbatova2020taxonomy} & \ref{grey:pytorch_styleguide}\ref{grey:tf_memory} & & \ref{grey:so_memory}\\

    Deterministic Algorithm Option Not Used & Model Training & Reproducibility & Generic & \cite{breck2017ml} & \ref{grey:pytorch_randomness}\\    
    
    Randomness Uncontrolled  & \makecell[l]{Model Training \\ \& Model Evaluation } & Reproducibility &  Generic & \cite{breck2017ml} & \ref{grey:ml_perfect}\ref{grey:pytorch_styleguide}\ref{grey:pytorch_randomness}& & \ref{grey:so_random_seed}\\
    
    Missing the Mask of Invalid Value & Model Training &  Error-prone &  Generic & \cite{zhang2018empirical}\cite{humbatova2020taxonomy} & & & \ref{grey:so_log1}\ref{grey:so_log2}\ref{grey:so_log3}\ref{grey:so_log4}\\
    
    Broadcasting Feature Not Used & Model Training & Efficiency & Generic & & \ref{grey:effective-tensorflow}\\
    
    TensorArray Not Used & Model Training & \makecell[l]{Efficiency \\ \& Error-prone} & API-Specific: TensorFlow 2 & & \ref{grey:effective-tensorflow} \\  
    
    Training / Evaluation Mode Improper Toggling & Model Training & Error-prone & Generic & & \ref{grey:pytorch_nn_common_mistakes}\\    
    
    Pytorch Call Method Misused  & Model Training & Robustness & API-Specific: PyTorch & & \ref{grey:pytorch_styleguide}\\  
    
    Gradients Not Cleared before Backward Propagation  & Model Training & Error-prone & API-Specific: PyTorch & & \ref{grey:pytorch_nn_common_mistakes} & &\\
    
    Data Leakage  & Model Evaluation &  Error-prone &  Generic & \cite{haakman2020studying} & \ref{grey:sklearn_best_practice} & & \ref{grey:so_dataleakage}\\
    
    Threshold-Dependent Validation  & Model Evaluation &  Robustness &  Generic & \cite{rajbahadur2019pitfalls} &\\

    \bottomrule
  \end{tabular}}
\end{table*}

\subsection{Unnecessary Iteration}

Avoid unnecessary iterations. Use vectorized solutions instead of loops.

\codeSmell{
Loops are typically time-consuming and verbose, while developers can usually use some vectorized solutions to replace the loops.
}{
As stated in the Pandas documentation~\ref{grey:pandas_vectorized}: ``Iterating through pandas objects is generally slow. In many cases, iterating manually over the rows is not needed and can be avoided''. In~\ref{grey:effective-tensorflow}, it is also stated that the slicing operation with loops in TensorFlow is slow, and there is a substitute for better performance.
}{
Machine learning applications are typically data-intensive, requiring operations on data sets rather than an individual value. Therefore, it is better to adopt a vectorized solution instead of iterating over data. In this way, the program runs faster and code complexity is reduced, resulting in more efficient and less error-prone code~\cite{haakman2020studying}. Pandas' built-in methods (e.g., join, groupby) are vectorized. It is therefore recommended to use Pandas built-in methods as an alternative to loops. In TensorFlow, using the \textit{tf.reduce\_sum()} API to perform reduction operation is much faster than combining slicing operation and loops.
}

\subsection{NaN Equivalence Comparison Misused}

The \textit{NaN} equivalence comparison is different to \textit{None} comparison. The result of \textit{NaN == NaN} is False~\ref{grey:nan_comparison}.

\codeSmell{
\textit{NaN} equivalence comparison behaves differently from \textit{None} equivalence comparison.
}{
While \textit{None} == \textit{None} evaluates to \textit{True}, \textit{np.nan} == \textit{np.nan} evaluates to \textit{False} in NumPy. As Pandas treats \textit{None} like \textit{np.nan} for simplicity and performance reasons, a comparison of \textit{DataFrame} elements with \textit{np.nan} always returns \textit{False} \cite{haakman2020studying}. If the developer is not aware of this, it may lead to unintentional behaviours in the code.
}{
Developers need to be careful when using the \textit{NaN} comparison. 
}

\subsection{Chain Indexing}
\label{chainindexing}

Avoid using chain indexing in Pandas.

\codeSmell{
In Pandas, \textit{df[``one''][``two'']} and \textit{df.loc[:,(``one'',``two'')]} give the same result. \textit{df[``one''][``two'']} is called chain indexing.
}{
Using chain indexing may cause performance issues as well as error-prone code~\ref{grey:pandas_indexing}\ref{grey:so_indexing1}\ref{grey:so_indexing2}. For example, when using \textit{df[``one''][``two'']}, Pandas sees this operation as two events: call \textit{df[``one'']} first and call \textit{[``two'']} based on the result the previous operation gets. On the contrary, \textit{df.loc[:,(``one'',``two'')]} only performs a single call. In this way, the second approach can be significantly faster than the first one. Furthermore, assigning to the product of chain indexing has inherently unpredictable results. Since Pandas makes no guarantees on whether \textit{df[``one'']} will return a view or a copy, the assignment may fail.
}{
Developers using Pandas should avoid using chain indexing.
}

\subsection{Columns and DataType Not Explicitly Set}
Explicitly select columns and set \textit{DataType} when importing data.
\codeSmell{
In Pandas, all columns are selected by default when a \textit{DataFrame} is imported from a file or other sources. The data type for each column is defined based on the default \textit{dtype} conversion.
}{
If the columns are not selected explicitly, it is not easy for developers to know what to expect in the downstream data schema~\ref{grey:pandas_style_guide}. If the datatype is not set explicitly, it may silently continue the next step even though the input is unexpected, which may cause errors later. The same applies to other data importing scenerios.
}{
It is recommended to set the columns and \textit{DataType} explicitly in data processing.
}

\subsection{Empty Column Misinitialization}
When a new empty column is needed in a \textit{DataFrame} in Pandas, use the \textit{NaN} value in Numpy instead of using zeros or empty strings.

\codeSmell{
Developers may need a new empty column in \textit{DataFrame}.
}{
If they use zeros or empty strings to initialize a new empty column in Pandas, the ability to use methods such as \textit{.isnull()} or \textit{.notnull()} is retained~\ref{grey:pandas_style_guide}. This might also happens to initializations in other data structure or libraries.
}{
Use \textit{NaN} value (e.g. ``\textit{np.nan}'') if a new empty column in a \textit{DataFrame} is needed. Do not use ``filler values'' such as zeros or empty strings.
}

\subsection{Merge API Parameter Not Explicitly Set}
Explicitly specify the parameters for merge operations. Specifically, explicitly specify \textit{on}, \textit{how} and \textit{validate} parameter for \textit{df.merge()} API in Pandas for better readability.

\codeSmell{
\textit{df.merge()} API merges two \textit{DataFrame}s in Pandas.  
}{
Although using the default parameter can produce the same result, explicitly specify \textit{on} and \textit{how} produce better readability~\ref{grey:pandas_style_guide}. The parameter \textit{on} states which columns to join on, and the parameter \textit{how} describes the join method (e.g., outer, inner). Also, the \textit{validate} parameter will check whether the merge is of a specified type. If the developer assumes the merge keys are unique in both left and right datasets, but that is not the case, and he does not specify this parameter, the result might silently go wrong. The merge operation is usually computationally and memory expensive. It is preferable to do the merging process in one stroke for performance consideration.
}{
Developer should explicitly specify the parameters for merge operation.
}

\subsection{In-Place APIs Misused}

Remember to assign the result of an operation to a variable or set the in-place parameter in the API.

\codeSmell{
Data structures can be manipulated in mainly two different approaches: 1) by applying the changes to a copy of the data structure and leaving the original object intact, or 2) by changing the existing data structure (also known as in-place).
}{
Some methods can adopt in-place by default, while others return a copy. If the developer assumes an in-place approach, he will not assign the returned value to any variable. Hence, the operation will be executed, but it will not affect the final outcome. For example, when using the Pandas library, the developer may not assign the result of \textit{df.dropna()} to a variable. He may assume that this API will make changes on the original \textit{DataFrame} and not set the in-place parameter to be \textit{True} either. The original \textit{DataFrame} will not be updated in this way~\cite{haakman2020studying}. In the \textit{``TensorFlow Bugs'' replication package}, we also found an example~\ref{grey:githubcommit_inplace} where the developer thought \textit{np.clip()} is an in-place operation and used it without assigning it to a new variable.
}{
We suggest developers check whether the result of the operation is assigned to a variable or the in-place parameter is set in the API. Some developers hold the view that the in-place operation will save memory. However, this is a misconception in the Pandas library because the copy of the data is still created. In PyTorch, the in-place operation does save GPU memory, but it risks overwriting the values needed to compute the gradient~\ref{grey:pytorch_inplace}.
}

\subsection{Dataframe Conversion API Misused}

Use \textit{df.to\_numpy()} in Pandas instead of \textit{df.values()} for transform a \textit{DataFrame} to a NumPy array.

\codeSmell{
In Pandas, \textit{df.to\_numpy()} and \textit{df.values()} both can turn a \textit{DataFrame} to a NumPy array.
}{
As noted in \ref{grey:so_df_conversion}, \textit{df.values()} has an inconsistency problem. With \textit{.values()} it is unclear whether the returned value would be the actual array, some transformation of it, or one of the Pandas custom arrays. However, the \textit{.values()} API has not been not deprecated yet. Although the library developers note it as a warning in the documentation, it does not log a warning or error when compiling the code if we use \textit{.value()}.
}{
When converting \textit{DataFrame} to NumPy array, it is better to use \textit{df.to\_numpy()} than \textit{df.values()}.
}

\subsection{Matrix Multiplication API Misused}

When the multiply operation is performed on two-dimensional matrixes, use \textit{np.matmul()} instead of \textit{np.dot()} in NumPy for better semantics. 

\codeSmell{
 When the multiply operation is performed on two-dimensional matrixes, \textit{np.matmul()} and \textit{np.dot()} give the same result, which is a matrix.
}{
In mathematics, the result of the dot product is expected to be a scalar rather than a vector~\ref{grey:dot_product}. The \textit{np.dot()} returns a new matrix for two-dimensional matrixes multiplication, which does not match with its mathematics semantics. Developers sometimes use \textit{np.dot()} in scenarios where it is not supposed to, e.g., two-dimensional multiplication.
}{
 When the multiply operation is performed on two-dimensional matrixes, \textit{np.matmul()} is preferred over \textit{np.dot()} for its clear semantic~\ref{grey:so_npdot}\ref{grey:np_dot}.
}

\subsection{No Scaling before Scaling-Sensitive Operation}

Check whether feature scaling is added before scaling-sensitive operations.

\codeSmell{
Feature scaling is a method of aligning features from various value ranges to the same range~\ref{grey:feature-scaling}.
}{
There are many operations sensitive to feature scaling, including Principal Component Analysis (PCA), Support Vector Machine (SVM), Stochastic Gradient Descent (SGD), Multi-layer Perceptron classifier and L1 and L2 regularization~\ref{grey:ml_wrong}\ref{grey:sklearn_scaling}. Missing scaling can lead to a wrong conclusion. For example, if one variable is on a larger scale than another, it will dominate the PCA procedure. Therefore, PCA without feature scaling can produce a wrong principal component result. 
}{
To avoid bugs, whether feature scaling is added before scaling-sensitive operations should be checked.
}

\subsection{Hyperparameter Not Explicitly Set}

Hyperparameters should be set explicitly.

\codeSmell{
Hyperparameters are usually set before the actual learning process begins and control the learning process~\cite{haakman2020studying}. These parameters directly influence the behavior of the training algorithm and therefore have a significant impact on the model's performance.
}{
The default parameters of learning algorithm APIs may not be optimal for a given data or problem, and may lead to local optima. In addition, while the default parameters of a machine learning library may be adequate for some time, these default parameters may change in new versions of the library. Furthermore, not setting the hyperparameters explicitly is inconvenient for replicating the model in a different programming language. 
}{
Hyperparameters should be set explicitly and tuned for improving the result's quality and reproducibility.
}

\subsection{Memory Not Freed}

Free memory in time.

\codeSmell{
Machine learning training is memory-consuming, and the machine's memory is always limited by budget.
}{
If the machine runs out of memory while training the model, the training will fail.
}{
Some APIs are provided to alleviate the run-out-of-memory issue in deep learning libraries.  TensorFlow's documentation notes that if the model is created in a loop, it is suggested to use \textit{clear\_session()} in the loop~\ref{grey:tf_memory}. Meanwhile, the GitHub repository \textit{pytorch-styleguide} recommends using \textit{.detach()} to free the tensor from the graph whenever possible~\ref{grey:pytorch_styleguide}. The \textit{.detach()} API can prevent unnecessary operations from being recorded and therefore can save memory~\ref{grey:detach}. Developers should check whether they use this kind of APIs to free the memory whenever possible in their code. 
}

\subsection{Deterministic Algorithm Option Not Used}

Set deterministic algorithm option to \textit{True} during the development process, and use the option that provides better performance in the production.

\codeSmell{
Using deterministic algorithms can improve reproducibility.
}{
The non-deterministic algorithm cannot produce repeatable results, which is inconvenient for debugging.
}{
Some libraries provide APIs for developers to use the deterministic algorithm. In PyTorch, it is suggested to set \textit{torch.use\_deterministic\_algorithms(True)} when debugging~\ref{grey:pytorch_randomness}. However, the application will perform slower if this option is set, so it is suggested not to use it in the deployment stage. 
}

\subsection{Randomness Uncontrolled}

Set random seed explicitly during the development process whenever a possible random procedure is involved in the application.

\codeSmell{
There are several scenarios involving random seeds. In some algorithms, randomness is inherently involved in the training process. For the cross-validation process in the model evaluation stage, the dataset split by some library APIs can vary depending on random seeds.
}{
If the random seed is not set, the result will be irreproducible, which increases the debugging effort. In addition, it will be difficult to replicate the study based on the previous one. For example, in Scikit-Learn, if the random seed is not set, the random forest algorithm may provide a different result every time it runs, and the dataset split by cross-validation splitter will also be different in the next run~\ref{grey:sklearn_best_practice}. 
}{
It is recommended to set global random seed first for reproducible results in Scikit-Learn, Pytorch, Numpy and other libraries where a random seed is involved~\ref{grey:ml_perfect}\ref{grey:pytorch_randomness}. Specifically, \textit{DataLoader} in PyTorch needs to be set with a random seed to ensure the data is split and loaded in the same way every time running the code. 
}

\subsection{Missing the Mask of Invalid Value}

Add a mask for possible invalid values. For example, developers should wrap the argument for \textit{tf.log()} with \textit{tf.clip()} to avoid the argument turning to zero.

\codeSmell{
In deep learning, the value of the variable changes during training. The variable may turn into an invalid value for another operation in this process.
}{
Several posts on Stack Overflow talk about the pitfalls that are not easy to discover caused by the input of the log function approaching zero \ref{grey:so_log1}\ref{grey:so_log2}\ref{grey:so_log3}\ref{grey:so_log4}. In this kind of programs, the input variable turns to zero and becomes an invalid value for \textit{tf.log()}, which raises an error during the training process. However, the error's stack trace did not directly point to the line of code that the bug exists \cite{zhang2018empirical}. This problem is not easy to debug and may take a long training time to find.
}{
The developer should check the input for the \textit{log} function or other functions that have special requirements for the argument and add a mask for them to avoid the invalid value. For example, developer can change \textit{tf.log(x)} to \textit{tf.log(tf.clip\_by\_value(x,1e-10,1.0))}. If the value of \textit{x} becomes zero, i.e., lower than the lowest bound 1e-10, the \textit{tf.clip\_by\_value()} API will act as a mask and outputs 1e-10. It will save time and effort if the developer could identify this smell before the code run into errors.
}

\subsection{Broadcasting Feature Not Used}

Use the broadcasting feature in deep learning code to be more memory efficient. 

\codeSmell{
Deep learning libraries like PyTorch and TensorFlow supports the element-wise broadcasting operation.
}{
Without broadcasting, tiling a tensor first to match another tensor consumes more memory due to the creation and storage of a middle tiling operation result~\ref{grey:effective-tensorflow}\ref{grey:broadcasting_pytorch}.  
}{
With broadcasting, it is more memory efficient. However, there is a trade-off in debugging since the tiling process is not explicitly stated.
}

\subsection{TensorArray Not Used}

Use \textit{tf.TensorArray()} in TensorFlow 2 if the value of the array will change in the loop.

\codeSmell{
Developers may need to change the value of the array in the loops in TensorFlow.
}{
If the developer initializes an array using \textit{tf.constant()} and tries to assign a new value to it in the loop to keep it growing, the code will run into an error. The developer can fix this error by the low-level \textit{tf.while\_loop()} API~\ref{grey:effective-tensorflow}. However, it is inefficient coding in this way. A lot of intermediate tensors are built in this process.
}{
Using \textit{tf.TensorArray()} for growing array in the loop is a better alternative for this kind of problem in TensorFlow 2. Developers should use new data types from libraries for more intelligent solutions.
}

\subsection{Training / Evaluation Mode Improper Toggling}

Call the training mode in the appropriate place in deep learning code to avoid forgetting to toggle back the training mode after the inference step.

\codeSmell{
In PyTorch, calling \textit{.eval()} means we are going into the evaluation mode and the \textit{Dropout} layer will be deactivated.
}{
If the training mode did not toggle back in time, the \textit{Dropout} layer would not be used in some data training and thus affect the training result~\ref{grey:pytorch_nn_common_mistakes}. The same applies to TensorFlow library.
}{
Developers should call the training mode in the right place to avoid forgetting to switch back to the training mode after the inference step.
}

\subsection{Pytorch Call Method Misused}

Use \textit{self.net()} in PyTorch to forward the input to the network instead of \textit{self.net.forward()}.

\codeSmell{
Both \textit{self.net()} and \textit{self.net.forward()} can be used to forward the input into the network in PyTorch.
}{
In PyTorch, \textit{self.net()} and \textit{self.net.forward()} are not identical. The \textit{self.net()} also deals with all the register hooks, which would not be considered when calling the plain \textit{.forward()} \ref{grey:pytorch_styleguide}. 
}{
It is recommended to use \textit{self.net()} rather than \textit{self.net.forward()}. 
}

\subsection{Gradients Not Cleared before Backward Propagation}

Use \textit{optimizer.zero\_grad()}, \textit{loss\_fn.backward()}, \textit{optimizer.step()} together in order in PyTorch. Do not forget to use \textit{optimizer.zero\_grad()} before \textit{loss\_fn.backward()} to clear gradients.

\codeSmell{
In PyTorch, \textit{optimizer.zero\_grad()} clears the old gradients from last step, \textit{loss\_fn.backward()} does the back propagation, and \textit{optimizer.step()} performs weight update using the gradients.
}{
If \textit{optimizer.zero\_grad()} is not used before \textit{loss\_fn.backward()}, the gradients will be accumulated from all \textit{loss\_fn.backward()} calls and it will lead to the gradient explosion, which fails the training~\ref{grey:pytorch_nn_common_mistakes}.
}{
Developers should use \textit{optimizer.zero\_grad()}, \textit{loss\_fn.backward()}, \textit{optimizer.step()} together in order and should not forget to use \textit{optimizer.zero\_grad()} before \textit{loss\_fn.backward()}.
}

\subsection{Data Leakage}

Use \textit{Pipeline()} API in Scikit-Learn or check data segregation carefully when using other libraries to prevent data leakage.

\codeSmell{
The data leakage occurs when the data used for training a machine learning model contains prediction result information~\ref{grey:data_leakage}.
}{
Data leakage frequently leads to overly optimistic experimental outcomes and poor performance in real-world usage~\cite{haakman2020studying}.
}{
There are two main sources of data leakage: leaky predictors and a leaky validation strategy~\ref{grey:data_leakage_kaggle}. Leaky predictors are the cases in which some features used in training are modified or generated after the goal value has been achieved. This kind of data leakage can only be inspected at the data level rather than the code level. Leaky validation strategy refers to the scenario where training data is mixed with validation data. This fault can be checked at the code level. One best practice in Scikit-Learn is to use the \textit{Pipeline()} API to prevent data leakage.
}

\subsection{Threshold-Dependent Validation}

Use threshold-independent metrics instead of threshold-dependent ones in model evaluation.

\codeSmell{
The performance of the machine learning model can be measured by different metrics, including threshold-dependent metrics (e.g., F-measure) or threshold-independent metrics (e.g., Area Under the Curve (AUC)).
}{
Choosing a specific threshold is tricky and can lead to a less-interpretable result~\cite{rajbahadur2019pitfalls}.
}{
Threshold-independent metrics are more robust and should be preferred over threshold-dependent metrics. 
}

\section{Discussions and Implications}

The code smell catalog summarized from the empirical study is presented in Table~\ref{tab:codesmellcatalog}. We collected 22 code smells in total and linked the smells to four pipeline stages: Data Cleaning, Feature Engineering, Model Training, and Model Evaluation. Possible impacts of the smells on application codes include being error-prone, less efficient, less reproducible, causing memory issues, less readable, and less robust. In addition, 16 smells are generic smells, while 6 are API-specific smells. Generic smells occur regardless of which library the developer uses, while API-specific smells depend on a specific library API design.

The catalog helps understand prevalent flaws in machine learning application development by investigating recurrent code issues from various sources. Since many data scientists do not have a software engineering background and are not up-to-date with the best practices from the software engineering field, our catalog of smells mitigates this barrier by providing some guidelines when developing machine learning applications.

Machine learning libraries are being regularly improved with new versions. We reused the ``TensorFlow Bugs'' replication package and found that many instances have already been deprecated because TensorFlow has upgraded to version 2. Hence, we expect that new API-specific code smells will appear with new versions and library features. In fact, our results showcase that most API-related smells are only reported by grey literature in general instead of literature. We argue that collecting a catalog of code smells helps in promoting a continuous effort between practitioners and academics.

The ecosystem of AI frameworks is changing very fast, which means that some smells might become obsolete in the meantime. In our catalog, we anticipate that three smells can be considered temporary smells: \textit{Dataframe Conversion API Misused}, \textit{Matrix Multiplication API Misused} and \textit{Gradients Not Cleared before Backward Propagation}. While other smells are perceived to last for a long time, temporary smells might be deprecated in a few years. Yet, these three smells are important and should be flagged to help practitioners prevent issues downstream.

\subsection{Implications to Data Scientists and Machine Learning Application Developers}

This catalog contains smells from heterogeneous sources, existing in different stages, and will trigger various effects. For instance, the \textit{Unnecessary Iteration} code smell describes the inefficient code structure and it often occurs at data cleaning stages. Another code smell \textit{Hyperparameter Not Explicitly Set} indicates irreproducible code and it is at model training stage. Data scientists and machine learning application developers can check these aspects while checking their code.

Some code smells appear multiple times in different sources -- both from academic and grey literature. For example, \textit{Missing the Mask of Invalid Value} is referenced in two instances of academic literature and four from Stack Overflow posts. Practitioners can use this as an indication of the relevance of smells and use the references to learn more about them.

Machine learning application developers, especially data scientists with little software engineering experience, can use the catalog to build awareness of the pitfalls and best practices highlighted in this study and strive to prevent these errors from their code. We assume that knowing code smells can shorten the time of development and help assure high-quality software in production. Future work will validate whether eliminating these code smells will lead to more accurate results during training, better hyperparameter optimization, clearer and higher quality code, and less maintenance effort. 

\subsection{Implication to Machine Learning Library Developers}

Some smells in the catalog stem from the fact that APIs require a particular usage pattern that is not intuitive to their users. For example, the smell \textit{Dataframe Conversion API Misused} smell could be eradicated if the API method \textit{df.values()} would be deprecated and completely replaced by \textit{df.to\_numpy()}. In another example, the \textit{Gradients Not Cleared before Backward Propagation} smell could be avoided if the API already took care of combining gradient clear and backward propagation for its users, since this is the commended approach. 
Hence, our results show how the design of library APIs plays an important role in avoiding potential issues in projects.

Some of the smells we identify are reported in the official documentation of the libraries. Yet, there is still code being created that does not comply the recommendations. For example, the effect of index chaining (cf. Section~\ref{chainindexing}) appears in code examples provided by Stack Overflow although it is explained in the Pandas documentation.
This indicates that many developers are struggling to follow the documentation strictly. It might stem from the fast iteration cycles in the development process of teams or from the developer's lack of experience in that particular library. We argue that passively indicating warnings on documentation might not be sufficient. It is important that library developers and maintainers are actively engaging in community forums, such as Stack Overflow, to help the community avoid non-obvious issues.

Finally, it is important that library maintainers promote and reach out to existing projects that aim at helping the development of machine learning software -- i.e., static code analysis tools, testing tools, quality auditors, experiment trackers, and so on.
Library developers know better than anyone what is the optimal way of leveraging their libraries. Hence, their contribution is crucial in the development of coding tools that support best practices.

\subsection{Implication to Code Analysis Tool Developers}

As some code smells cannot be addressed by designing better APIs, the static analysis tool can help promote best practices and warn pitfalls to the application developers.

This research serves as the base for future work on automated tools to detect these unwanted code patterns. Automated tools can minimize the developer's effort to discover the code smells and eliminate them, providing support for code quality assurance. Because humans are occasionally forgetful, it is preferable to have a technology that expressly checks whether best practices are being followed.

In addition, we observe that some code smells are related to the context. This is aligned with previous work that proposes context-aware code analysis tools for machine learning applications~\cite{kannan2022mlsmellhound}. For example, PyTorch library developers recommend application developers to use the deterministic option during the development but not set it in the production code due to the consideration for performance. Therefore, the automated tool can have different configuration settings. For example, according to the pipeline stage, it can have a development setting and a deployment setting. 

\subsection{Implication to Students}

As mentioned by \cite{haakman2020studying}, many graduates in the industry do not have formal education on machine learning application development since it requires a combination of software engineering and data science practices. Students can use this catalog to learn more about the common anti-patterns in machine learning application development and prepare for future jobs.

\section{Threats to validity}

In our study, the first author performs manual code smell inspections, which can be biased due to the different understanding of machine learning code. To alleviate this threat, the second author reviews all instances of code smells, followed by a discussion between the first two authors.

In both the academic and grey literature survey, the initial selection of keywords in the search query might miss relevant entries. To mitigate this threat, we iteratively refine the search keywords based on retrieved relevant content. In addition, we apply forward and backward snowballing to complement the search.

Moreover, since we use a back-cutting strategy on the grey literature search, the quality of the search results depends on the accuracy of the Google search engine's relevance sorting algorithm, which is beyond our control. The results are collected from the first author's Google account, and they might vary across users. However, we believe that this has minimal impact on the result set of our study. 

When mining Stack Overflow entries and GitHub commits, we inspect 88 GitHub commits and 491 Stack Overflow posts in total. It is unclear how generalizable our results are. To cover the most common mistakes in the machine learning application practice in a generalizable way, we use the ``highest voted'' criteria to select instances from Stack Overflow. We anticipate that less-voted instances may also contain machine learning code issues. Increasing the result set would not be feasible in a manual inspection. Yet, we argue that the highest voted instances provide an interesting snapshot with the most relevant issues. 

This study focuses on six Python machine learning libraries and frameworks. There are several other machine learning frameworks that might lead to particular code smells. However, it would not be feasible to apply our methodology in all the libraries out there. Hence, we reduce this threat by selecting the most popular frameworks.

Finally, we acknowledge that there are more warnings within libraries documentation that can become code smells. However, we only consider warnings that have allegedly led to real code issues, as observed in other sources (e.g., Stack Overflow).

\section{Conclusion and Future Work}

In this paper, we conducted an empirical study to collect the code smell specific for machine learning applications. We collected the code smells from various sources, including mining 1750 papers, mining 2170 grey literature entries, using the existing bugs datasets including 88 Stack Overflow posts and 87 GitHub commits and gathering 403 complementary Stack Overflow posts. We analyzed the pitfalls mentioned in the posts and decided whether to take it as a code smell. We collected 22 code smells, including general and API-specific smells. We also classified the code smell by different pipeline stages and its effect. We want to raise the discussion about machine learning-specific code smell and help improve code quality in the machine learning community in this way. 

Future work will include a quantitative large-scale validation of the code smell catalog. We would like to interview machine learning practitioners and mine code changes in GitHub repositories to validate and improve the catalog. In addition, we plan to implement a static analysis tool that automatically detect these smells to promote best practices in machine learning code. Finally, it would be interesting to study the prevalence of these code smells in real-world machine learning applications and explore the benefits of using a catalog of machine learning-specific code smells.

\begin{acks}
    This work was partially supported by ING through the AI for Fintech Research Lab with the Delft University of Technology.
\end{acks}

\bibliographystyle{ACM-Reference-Format}
\bibliography{sample-base}

\appendix
\section{Grey Literature References}
\label{sec:grey_literature}
\footnotesize
% \begin{enumerate}[label={[\arabic{enumi}]},ref=\arabic{enumi}]
% \begin{enumerate}[label=(\alph*),ref=(\alph*)]
\begin{enumerate}[label=(\arabic{enumi}),ref=(\arabic{enumi})]

  \item\label{grey:ml_perfect} Christian Haller. My Machine Learning Model Is Perfect.
  URL:~\url{https://towardsdatascience.com/my-machine-learning-model-is-perfect-9a7928e0f604}
  
  \item\label{grey:ml_wrong} Cheng-Tao Chu. Machine Learning Done Wrong. URL: ~\url{https://ml.posthaven.com/machine-learning-done-wrong}

  \item\label{grey:ml-common-mistakes} What are common mistakes when working with neural networks? URL:~\url{https://www.kaggle.com/general/196487} 
  
  \item\label{grey:mistakes-by-ds-scientist} Top 10 Coding Mistakes Made by Data Scientists. URL:~\url{https://www.kdnuggets.com/2019/04/top-10-coding-mistakes-data-scientists.html}
  
  \item\label{grey:pytorch_styleguide} Igor Susmelj, Lucas Vandroux, Daniel Bourke (2022). A PyTorch Tools, best practices \& Styleguide.
  URL: ~\url{https://github.com/IgorSusmelj/pytorch-styleguide}  

  \item\label{grey:effective-tensorflow} EffectiveTensorflow.
  URL:~\url{https://github.com/vahidk/EffectiveTensorflow}
  
  \item\label{grey:pandas_style_guide} Josh Levy-Kramer (2021). Pandas Style Guide.
  URL:~\url{https://github.com/joshlk/pandas_style_guide}
  
  \item\label{grey:sklearn_best_practice}  Scikit-Learn Documentation. URL: ~\url{https://scikit-learn.org/stable/common_pitfalls.html}  

  \item\label{grey:pytorch_randomness}  PyTorch Documentation. Reproducibility. URL: ~\url{https://pytorch.org/docs/stable/notes/randomness.html}
  
  \item\label{grey:pytorch_inplace}  Alexandra Deis. In-place Operations in PyTorch. URL: ~\url{https://towardsdatascience.com/in-place-operations-in-pytorch-f91d493e970e}
  
  \item\label{grey:githubcommit_inplace}  GitHub Commit. URL: ~\url{https://github.com/bamos/dcgan-completion.tensorflow/commit/e8b930501dffe01db423b6ca1c65d3ac54f27223}

  \item\label{grey:inplace} Samual Sam (2018). Inplace operator in Python. URL: ~\url{https://www.tutorialspoint.com/inplace-operator-in-python}  
  
  \item\label{grey:githubcommit_tf_verctorized} Github Commit -- Tensor Flow. URL: ~\url{https://github.com/tensorflow/models/commit/90f63a1e1653}
  
  \item\label{grey:pandas_vectorized} Pandas Documentation. Essential basic functionality -- Iteration. URL: ~\url{https://pandas.pydata.org/pandas-docs/stable/user_guide/basics.html#iteration}

  \item\label{grey:vectorized_blog} Vectorization, Part 2: Why and What? URL: ~\url{https://www.quantifisolutions.com/vectorization-part-2-why-and-what/}  
  
  \item\label{grey:sklearn_scaling} Scikit-Learn Documentation. URL: ~\url{https://scikit-learn.org/stable/modules/preprocessing.html}
  
  \item\label{grey:so_scaling} Stack Overflow. GridSearchCV extremely slow on small dataset in scikit-learn. URL: ~\url{https://stackoverflow.com/questions/17455302/gridsearchcv-extremely-slow-on-small-dataset-in-scikit-learn/23813876#23813876}
  
  \item\label{grey:feature-scaling} Feature scaling. URL:~\url{https://en.wikipedia.org/wiki/Feature_scaling}
  
  \item\label{grey:tf_memory} TensorFlow Documentation. Backend: \texttt{clear\_session}. URL: ~\url{https://www.tensorflow.org/api_docs/python/tf/keras/backend/clear_session}
  
  \item\label{grey:so_memory} Stack Overflow. Tensorflow OOM on GPU. URL: ~\url{https://stackoverflow.com/questions/42495930/tensorflow-oom-on-gpu}  
  
%   \item\label{grey:so_counterintuitive1} Stack Overflow. Why does TensorFlow example fail when increasing batch size? URL: ~\url{https://stackoverflow.com/questions/33641799/why-does-tensorflow-example-fail-when-increasing-batch-size}
  
%   \item\label{grey:so_counterintuitive2} Stack Overflow. Tensorflow Weights Diverge or NaN. URL: ~\url{https://stackoverflow.com/questions/43636736/tensorflow-weights-diverge-or-nan}
  
%   \item\label{grey:so_counterintuitive3} Stack Overflow. Tensorflow: loss becomes 'NaN'. URL: ~\url{https://stackoverflow.com/questions/43948571/tensorflow-loss-becomes-nan}
  
%   \item\label{grey:so_counterintuitive4} Stack Overflow. Tensorflow Issues. URL: ~\url{https://stackoverflow.com/questions/46577203/tensorflow-issues}
  
  \item\label{grey:so_log1} Stack Overflow. Tensorflow NaN bug? URL: ~\url{https://stackoverflow.com/questions/33712178/tensorflow-nan-bug}
  
  \item\label{grey:so_log2} Stack Overflow. TensorFlow's ReluGrad claims input is not finite. URL: ~\url{https://stackoverflow.com/questions/33699174/tensorflows-relugrad-claims-input-is-not-finite}
  
  \item\label{grey:so_log3} Stack Overflow. Tensorflow - Convolutionary Net: Grayscale vs Black/White training. URL: ~\url{https://stackoverflow.com/questions/39487825/tensorflow-convolutionary-net-grayscale-vs-black-white-training}
  
  \item\label{grey:so_log4} Stack Overflow. Implement MLP in tensorflow. URL: ~\url{https://stackoverflow.com/questions/35078027/implement-mlp-in-tensorflow}
  
  \item\label{grey:weight_initialized} Weight Initialization Techniques in Neural Networks. URL: ~\url{https://towardsdatascience.com/weight-initialization-techniques-in-neural-networks-26c649eb3b78}
  
  \item\label{grey:so_random_seed}  Stack Overflow. Best practices for generating a random seeds to seed Pytorch? URL: ~\url{https://stackoverflow.com/questions/57416925/best-practices-for-generating-a-random-seeds-to-seed-pytorch}  
  
  \item\label{grey:so_dataleakage} Stack Overflow. Keras Regression using Scikit Learn StandardScaler with Pipeline and without Pipeline. URL: ~\url{https://stackoverflow.com/questions/43816718/keras-regression-using-scikit-learn-standardscaler-with-pipeline-and-without-pip/43816833#43816833}
  
  \item\label{grey:data_leakage} Ask a Data Scientist: Data Leakage. URL: ~\url{https://insidebigdata.com/2014/11/26/ask-data-scientist-data-leakage/}
  
  \item\label{grey:data_leakage_kaggle} Data Leakage. URL: ~\url{https://www.kaggle.com/alexisbcook/data-leakage}  
  
  \item\label{grey:pandas_indexing} Pandas Documentation.  URL: ~\url{https://pandas.pydata.org/pandas-docs/stable/user_guide/indexing.html#indexing-view-versus-copy}
  
  \item\label{grey:so_indexing1} Stack Overflow. Extrapolate values in Pandas DataFrame. URL: ~\url{https://stackoverflow.com/questions/22491628/extrapolate-values-in-pandas-dataframe/35959909#35959909}
  
  \item\label{grey:so_indexing2} Stack Overflow. Why does one use of iloc() give a SettingWithCopyWarning, but the other doesn't? URL: ~\url{https://stackoverflow.com/questions/53806570/why-does-one-use-of-iloc-give-a-settingwithcopywarning-but-the-other-doesnt/53807453#53807453}
  
  \item\label{grey:so_df_conversion} Stack Overflow. Convert pandas dataframe to NumPy array. URL: ~\url{https://stackoverflow.com/questions/13187778/convert-pandas-dataframe-to-numpy-array/54508052#54508052}
  
%   \item\label{grey:so_regx} Stack Overflow. Fast punctuation removal with pandas. URL: ~\url{https://stackoverflow.com/questions/50444346/fast-punctuation-removal-with-pandas/50444347#50444347}
  
%   \item\label{grey:so_backend} Stack Overflow. Dynamically evaluate an expression from a formula in Pandas. URL: ~\url{https://stackoverflow.com/questions/53779986/dynamically-evaluate-an-expression-from-a-formula-in-pandas/53779987#53779987}
  
%   \item\label{grey:so_unaligned_tensor} Stack Overflow. Tensor with unspecified dimension in tensorflow URL: ~\url{https://stackoverflow.com/questions/34079787/tensor-with-unspecified-dimension-in-tensorflow}
  
%   \item\label{grey:so_initialization} Stack Overflow. Tensorflow: Using Adam optimizer. URL: ~\url{https://stackoverflow.com/questions/33788989/tensorflow-using-adam-optimizer}
  
  \item\label{grey:so_npdot} Stack Overflow. Does np.dot automatically transpose vectors? URL: ~\url{https://stackoverflow.com/questions/54160155/does-np-dot-automatically-transpose-vectors/54161169#54161169}
  
  \item\label{grey:np_dot} Linear Algebra (\texttt{numpy.dot}). NumPy Documentation. URL: ~\url{https://numpy.org/doc/stable/reference/generated/numpy.dot.html#numpy.dot}
  
%   \item\label{grey:most_popular_library_blog} 15 Python Libraries for Data Science You Should Know. URL: ~\url{https://www.dataquest.io/blog/15-python-libraries-for-data-science/}
  
%   \item\label{grey:most_popular_library_kaggle} State of Data Science and Machine Learning 2021. URL: ~\url{https://www.kaggle.com/kaggle-survey-2021}
  
  \item\label{grey:pytorch_nn_common_mistakes} Yuval Greenfield. Most Common Neural Net PyTorch Mistakes. URL: ~\url{https://medium.com/missinglink-deep-learning-platform/most-common-neural-net-pytorch-mistakes-456560ada037}
  
  \item\label{grey:so_dataloader} Stack Overflow. Is this a right way to train and test the model using Pytorch? URL: ~\url{https://stackoverflow.com/questions/67066452/is-this-a-right-way-to-train-and-test-the-model-using-pytorch/67067242#67067242}
  
  \item\label{grey:detach} Why does detach reduce the allocated memory? URL: ~\url{https://discuss.pytorch.org/t/why-does-detach-reduce-the-allocated-memory/43836}
  
  \item\label{grey:dot_product} Dot product. Wikipedia. URL: ~\url{https://en.wikipedia.org/wiki/Dot_product}

  \item\label{grey:nan_comparison} Stack Overflow. What is the rationale for all comparisons returning false for IEEE754 NaN values? URL: ~\url{https://stackoverflow.com/questions/1565164/what-is-the-rationale-for-all-comparisons-returning-false-for-ieee754-nan-values}
  
  \item\label{grey:broadcasting_pytorch} Broadcasting the good and the ugly URL: ~\url{https://effectivemachinelearning.com/PyTorch/3._Broadcasting_the_good_and_the_ugly}

\end{enumerate}
% \item\label{grey:}  URL: ~\url{ }
 
% \end{enumerate}

\end{document}